\documentclass[10pt]{article}

\usepackage{amsmath}
\usepackage{amssymb}

\usepackage{graphicx}
\usepackage{epsfig}
\usepackage{cite}

\usepackage{color}

\usepackage[labelfont=bf,labelsep=period,justification=raggedright]{caption}

\makeatletter
\renewcommand{\@biblabel}[1]{\quad#1.}
\makeatother

\date{}

\begin{document}

\begin{flushleft}
{\Large
\textbf{Non-Markovian Character in Human Mobility: Online and Offline}
}
\\
Zhi-Dan Zhao $^{1}$,
Shi-Min Cai $^{1,2,\ast}$,
Yang Lu $^{1}$
\\
\bf{1} Web Sciences Center, School of Computer Science and Engineering, University of Electronic Science and Technology of China, Chengdu 611731, P. R. China
\\
\bf{2} State Key Laboratory of Networking and Switching Technology, Beijing University of Posts and Telecommunications, Beijing, 100876, P. R. China
\\
$\ast$ E-mail: shimin.cai81@gmail.com
\end{flushleft}

\section*{Abstract}
The dynamics of human mobility characterizes the trajectories humans follow during their
daily activities and is the foundation of processes from epidemic spreading to traffic prediction
and information recommendation. In this paper, we investigate a massive data set of human activity including both online behavior of
browsing websites and offline one of visiting towers based mobile terminations. The
non-Markovian character observed from both online and offline cases is suggested by
the scaling law in the distribution of dwelling time at individual and collective levels, respectively.
Furthermore, we argue that the lower entropy and higher predictability in
human mobility for both online and offline cases may origin from this non-Markovian character.
However, the distributions of individual entropy and predictability show the different degrees of non-Markovian character from online to offline cases. To accounting for non-Markovian character in human mobility, we introduce a protype model with three basic ingredients, \emph{preferential return, inertial effect, and exploration}
to reproduce the dynamic process of online and offline human mobility. In comparison with standard and biased random walk models with assumption of Markov process, the proposed model is able to obtain characters much closer to these empirical observations.

\section*{Introduction}
Uncovering the dynamics of human mobility is pivotal to decipher the behavior patterns of human daily activities, with widely practical applications ranging from epidemic containment~\cite{Hufnagel2004,Colizza2007,Vespignani2009,Wang2009,Yang2012} to traffic predication~\cite{Helbing2001,Simini2012} and information recommendation~\cite{Balcan2011,Lu2012}.
The pursuit has been facilitated greatly by advances in information technology, especially by the availability of massive Internet data and resources including the real-time tracking trajectories of human mobility at temporal and spatial scales~\cite{Starnini2013}. Large amounts of studies on human mobility are therefore generated based
on these trajectories in physical space, such as dollar dispersal~\cite{Brockmann2006}, GPS~\cite{Rhee2008}, mobile phone~\cite{Gonzalez2008,Song2010,Song2010a,Simini2012}, questionnaire~\cite{Yan2013}, or cyber space, such as online web surfing~\cite{Huberman1998,Chmiel2009} and interest transition~\cite{Zhao2013a,Zhao2013b}. In these works, several widely accepted indicators, e.g., the trip distance distribution, the radius of gyration, the number of visited locations over time, the human mobility motifs, the consecutively dwelling time (or click length) for a location, the entropy and predictability, are used to statistically characterize the dynamics of human mobility~\cite{Brockmann2006,Gonzalez2008,Chmiel2009,Song2010,Song2010a,Schneider2013,Zhao2013a}.

Since Einstein's study~\cite{Einstein1905} on the motions of small particles suspended in liquids, diffusion and
random walk have been seemed as a paradigm describing and modeling the mobility in many cases of physical and biological domains ranging from charge transport in disordered conductors to foraging patterns of animals and fishes~\cite{Bouchaud1990,Ben2000}. On the other hand, a widely used assumption of Markovian (memoryless) dynamics is deemed to be existence in online human activities, which suggests that click actions are independent each
other~\cite{Brin1998,craswell2007,fagin2001}. However, recent evidences~\cite{Meiss2010,Chierichetti2012,Zhao2013a} demonstrate that online human activities may obey a non-Markov process.

Although many studies have focused on human mobility recently,
it is still obscure that whether individual mobilities
for both online and offline cases exactly follow a Markov process
and why there are lower entropy and higher predictability in human mobility.
In this paper, our systematical analysis of a massive data set commonly including both online and offline
human activities for same individuals seeks to more clearly unveil the dynamics of human mobility.
The dynamic processes of human mobility for both online and offline cases are translated into
visual networks describing that the participants transit from one website (or tower) to another. Based on that,
both at individual and collective levels, the heterogeneous rank distributions of nodes (i.e., websites and towers) according to visit frequency suggest that the human mobility follows a nonuniform process and the scaling law in the distribution of dwelling time shows the non-Markovian character in human mobility. Furthermore, the non-Markovian character imply the existence of memory in dynamics of human mobility, leading to a lower entropy and higher predictability in human mobility. In turn, the diverse distributions of individual entropy and
predictability indicate the different degree of non-Markovian character between online and offline cases.
By comparing with two Markov processes, Standard Random Walk (SRW) and Biased Random Walk (BRW), we
account for the non-Mrakovian character and origin of lower entropy and higher predictability in human mobility based on a protype model with three basic ingredients, preferential return, inertia effect and exploration, underlying mechanisms governing the dynamics of online and offline human mobility. Our results to some extent
uncover that the underlying mechanism governing the dynamics of human mobility is generic no matter of the temporal and spacial differences among their activities.

\section*{Results}

\subsection*{Empirical observations}
We explore a massive data set of human activity including online behavior of browsing websites and offline one of visiting towers based mobile terminations, whose trajectories were recorded by sequences of websites and towers at temporal scale, respectively. The non-Markovian character in human mobility is investigated with these aspects, 1) visual networks of human mobility; 2) rank distributions of nodes and distributions of dwelling time at individual and collective levels, respectively; 3) the individual entropy and predictability in human mobility over whole duration.

\begin{figure*}[!ht]
\begin{center}
\epsfig{figure=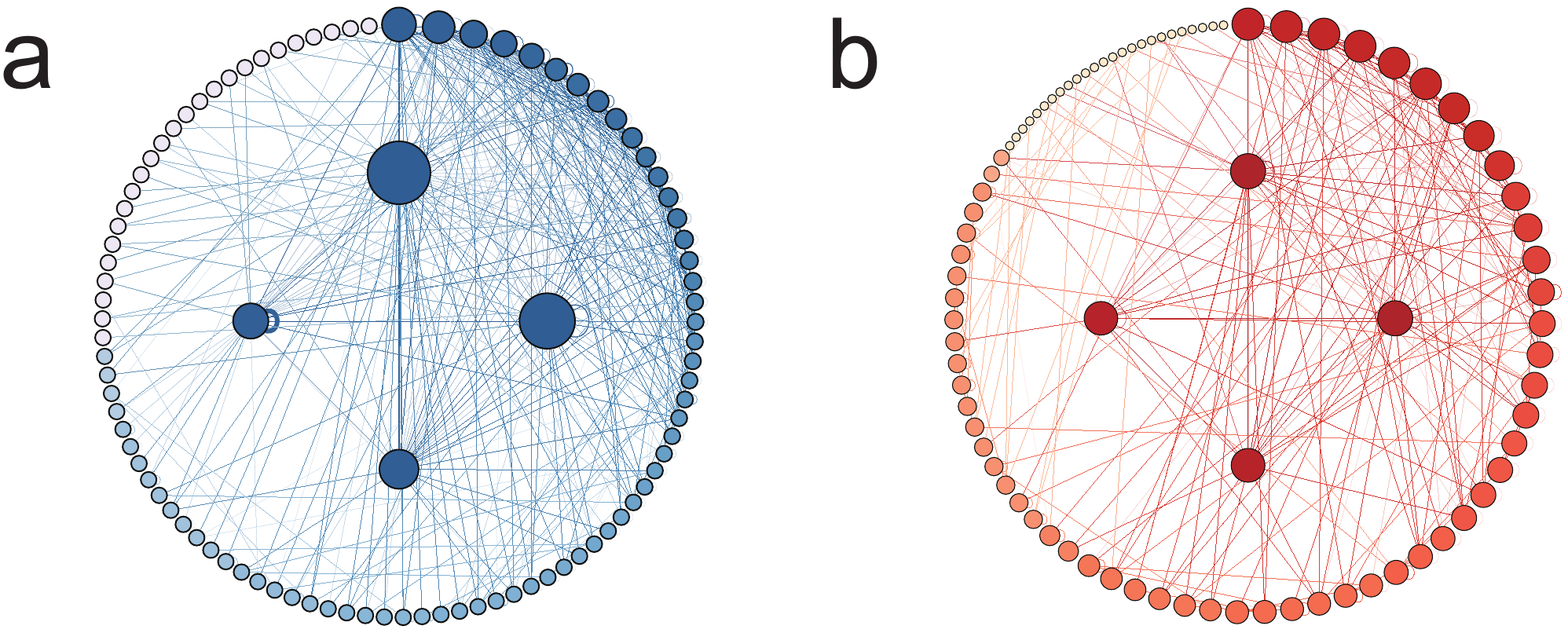,width=0.9\textwidth}
\end{center}
\caption{\baselineskip=12pt
{\bf Visual network of human mobility}.
For a selected participant, the visual networks for online (\textbf{a}) and offline (\textbf{b})
cases describe the dynamic processes of browsing distinct websites and visiting different towers, respectively.
In these networks, node denotes website or tower and weighted link indicates the intensity of transition.
Node size is determined by its visit frequency and self-loop represents the repeat visits to the same node.
A few frequently visited node are marked in the central area of network. The numbers of clicks ($N_a$) for
two cases are $6,666$ and the number of nodes are $95$.}
\label{fig:transition}
\end{figure*}

\emph{Visual network of human mobility}. An individual can visit different websites (towers) during her/his communicating activity based mobile termination. A convenient way to represent the direct associations among them is to construct a visual network where nodes denote websites (towers) and links correspond to observed transitions among websites (towers), shown in Fig.~\ref{fig:transition}(a) and (b) respectively. The node sizes and link weights are determined by visit and transition frequency respectively, and the self-loops of nodes indicates the dwelling time. From Fig.~\ref{fig:transition}(a) and (b), we can find that the networks of human mobility are obviously heterogeneous, suggesting the human mobility is a nonuniform process.

\begin{figure*}[!ht]
\begin{center}
\epsfig{figure=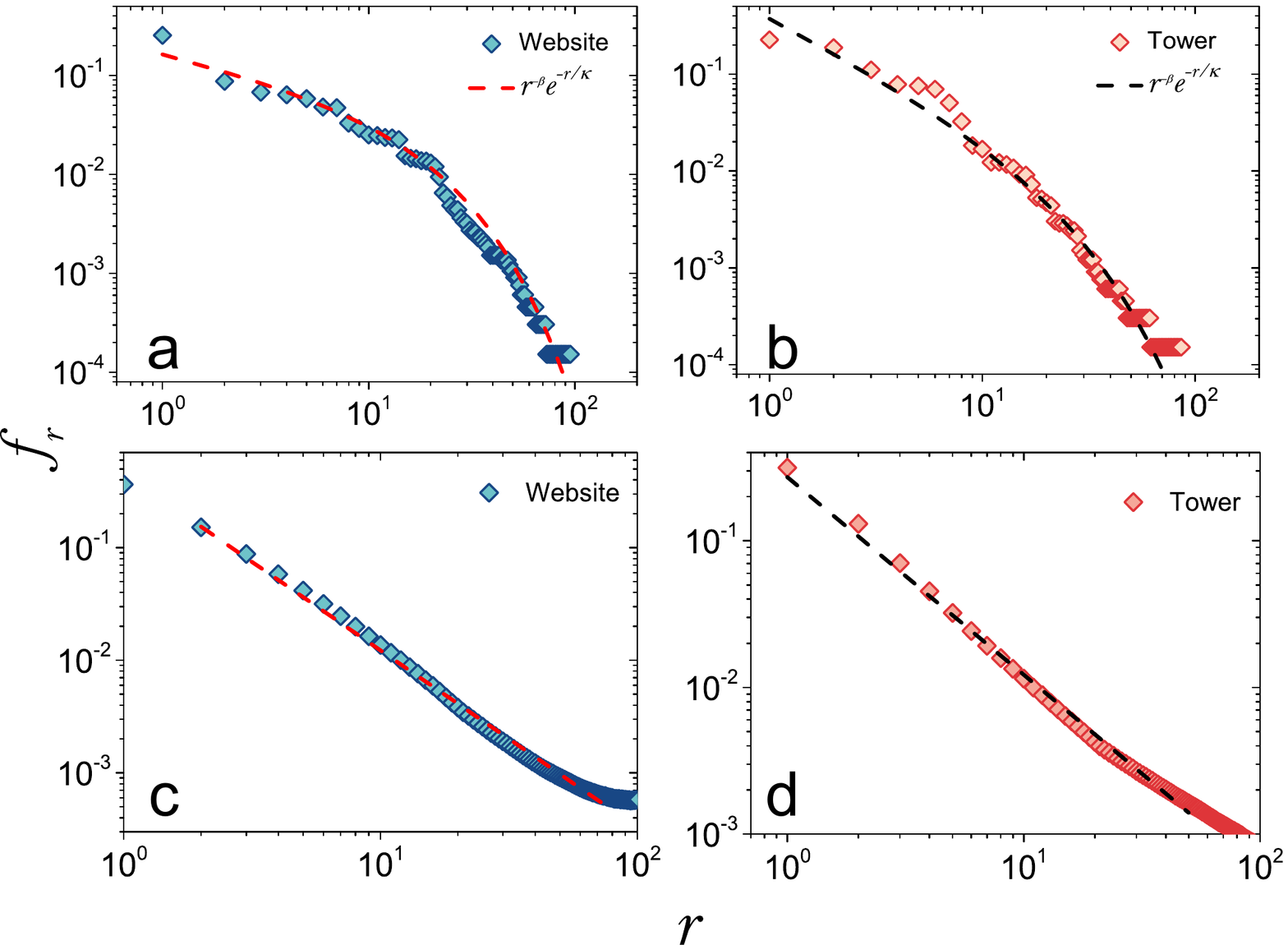,width=0.9\textwidth}
\end{center}
\caption{\baselineskip=12pt
{\bf Rank distribution of nodes of visual networks}.
The rank distribution at individual level shows the truncated power-law scaling,
$f_{r}\propto {r^{-\beta}}\exp{( - r/\kappa)}$, where the fitted values of
the exponent $\beta$ and $\kappa$ are $(\beta,\kappa)=(0.50,20)$
for online case \textbf{(a)} and $(\beta,\kappa)=(1.15,20)$ for offline case \textbf{(b)}.
For the whole system, we scaled the aggregated node size with the number of participants
who visit it and rank them. The rank distribution at collective level becomes a power-law scaling,
$\beta = 1.57 \pm 0.02$ for online case \textbf{(c)} and $\beta = 1.31 \pm 0.01$ for offline case \textbf{(d)}.
The similar pattern of human mobility for online and offline cases at two levels suggests that their underling
mechanisms governing the dynamics of human mobility may be same.}
\label{fig:noderank}
\end{figure*}

\emph{Rank distribution of nodes}. At individual level, the nodes are visited in different frequency, leading to heterogeneous node sizes. We rank nodes according to their size and probe its rank distribution. In Fig~\ref{fig:noderank}(a) and (b), the rank distributions for both online and offline cases are approximately fitted by an exponentially truncated power-law formula, $f_{r} = r^{-\beta}\exp{(-r/\kappa)}$, with values of $\beta$ are 0.5 and 1.15 for online and offline cases, respectively (the values of the exponent $\beta$ are tested by Kolmogorov-Smirnov (KS) test). Note that this truncated fat-tailed distribution is robust in respect to other individuals (see Fig.~S1 in Supplementary Information). For whole system, we are also interesting that the rank distribution of nodes when their sizes are aggregated at collective level. As the intensity of individual activity is different, we scaled the aggregated node size with the number of participants who visit it to filter its effect. In Fig~\ref{fig:noderank}(c) and (d), the rank distributions for both online and offline cases at collective level shows a power-law formula, $f_{r} = r^{-\beta}$, similar to the Zip'f law characterizing the rank distribution of locations in human mobility~\cite{Gonzalez2008,Song2010,Bagrow2012}. The values of $\beta$, $1.57$ and $1.31$ obtained from maximum likelihood estimation, are close with each other. Additionally, compared with that at individual level, the scaling law at collective level reduce the exponential truncation, which may result from the remove of effect of intensity of individual activity. These heterogeneous rank distributions of nodes at two level suggest the nonuniformity of dynamic process of human mobility and may imply the non-Markovian charter in human mobility. Simultaneously, considering the similar pattern of human mobility for online and offline cases at two levels, we state that their underlying mechanisms governing the dynamics of human mobility may be same.

\begin{figure*}[!ht]
\begin{center}
\epsfig{figure=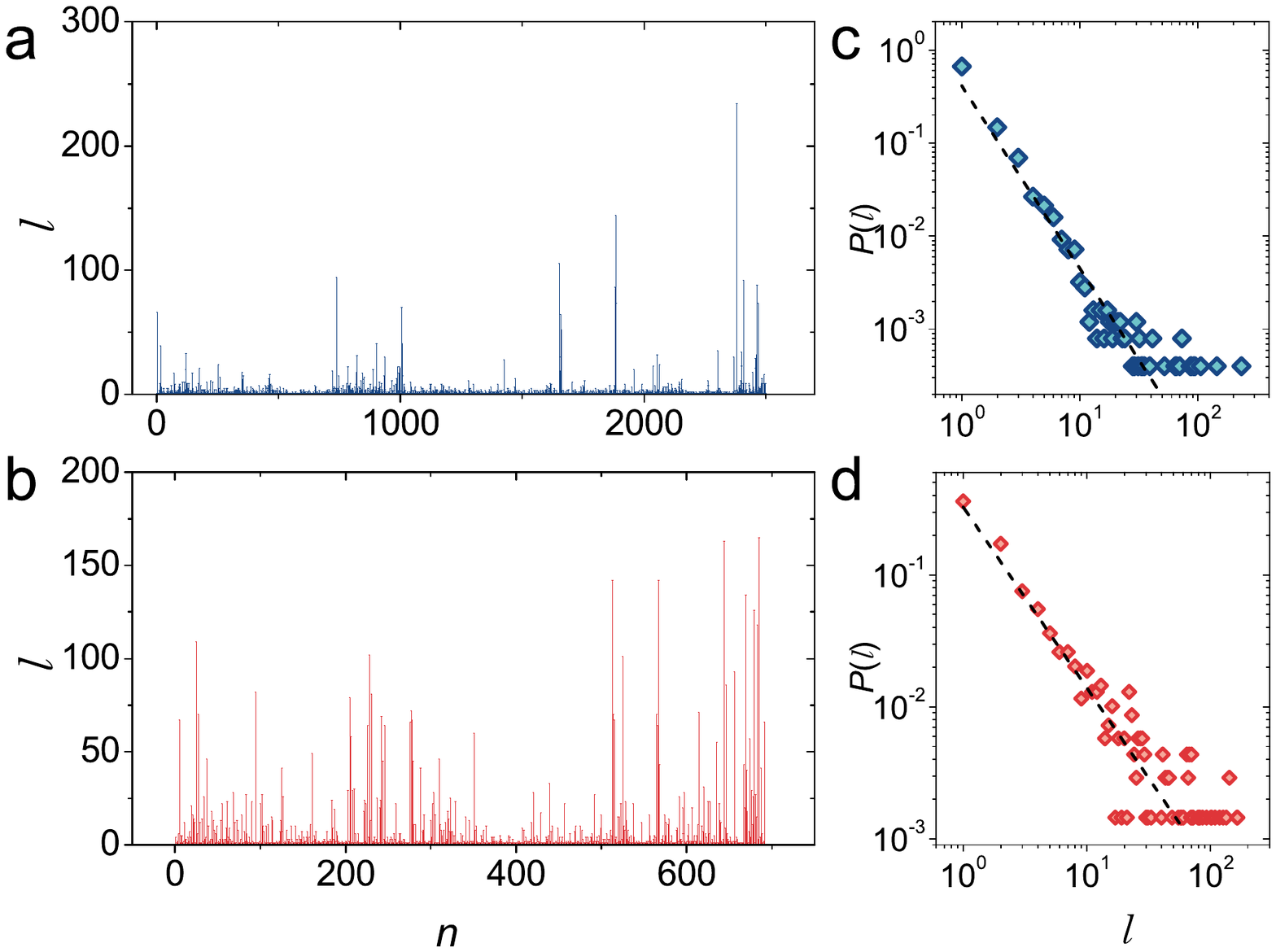,width=0.9\textwidth}
\end{center}
\caption{
\baselineskip=12pt
{\bf The sequence of dwelling time and its probability distribution at individual level}.
The sequences $l$ of dwelling time in respect to click $n$ for online \textbf{(a)} and offline \textbf{(b)} cases, obtained from the same participant, show the burst and clustering behavior.
The corresponding probability distribution $P(l)$ follows power-law scaling, $P(l) \sim l^{-\alpha}$,
where the exponent $\alpha$, with exponents $\alpha \approx 1.96$ and $1.36$ for online \textbf{(c)} and
offline \textbf{(d)} cases, respectively. The values of the exponent $\alpha$ are estimated by using the maximum likelihood criterion~\cite{Clauset2009}. These empirical results suggest the non-Markovian character in human mobility.}
\label{fig:dwelltime}
\end{figure*}

\emph{Distribution of dwelling time}.
Many approaches have been proposed to quantify human mobility, yet here we are interesting in the dwelling time $l$ consecutively visiting same node as a result of taking advantage of the nature of enough large data set. Additionally, the dwelling time associates with the inter-arrival time, whose statistical distribution can characterizes the Markovian property of stochastic process \cite{Kingman1963}. Like the analysis of rank distribution of node, we first focus on the statistical measurements of dwelling time for both online and offline cases at individual level. A group of typical sequences of $l$ corresponding to online and offline cases are shown Fig.~\ref{fig:dwelltime}(a) and (b), respectively. They both oppose some extremely large values and behave a clustering behavior, which result in their power-law distributions presented in Fig.~\ref{fig:dwelltime}(c) and (d). Much more concretely, for the distribution $P(l) \sim l^{-\alpha}$, a larger $\alpha$ indicates a much more irregular sequence of $l$, and the fat tail suggests that participant tends to take abnormally long time consecutively visiting same node. By fitting the power-law distributions based on maximum likelihood estimation~\cite{Clauset2009}, we obtain
$\alpha$ $\approx$ $1.96$ and $1.36$ for online and offline cases, respectively, confirmed by the irregularity of sequence shown in Fig.~\ref{fig:dwelltime}(a) and (b). Moreover, the scaling law proves the existence of non-Markovian character in human mobility because Markov process shows exponential~\cite{Kingman1963}. Note that much more participants from this massive data set have been measured to confirm the stability of scaling law (see Fig.~S2 in Supplementary Information).

\begin{figure*}[!ht]
\begin{center}
\epsfig{figure=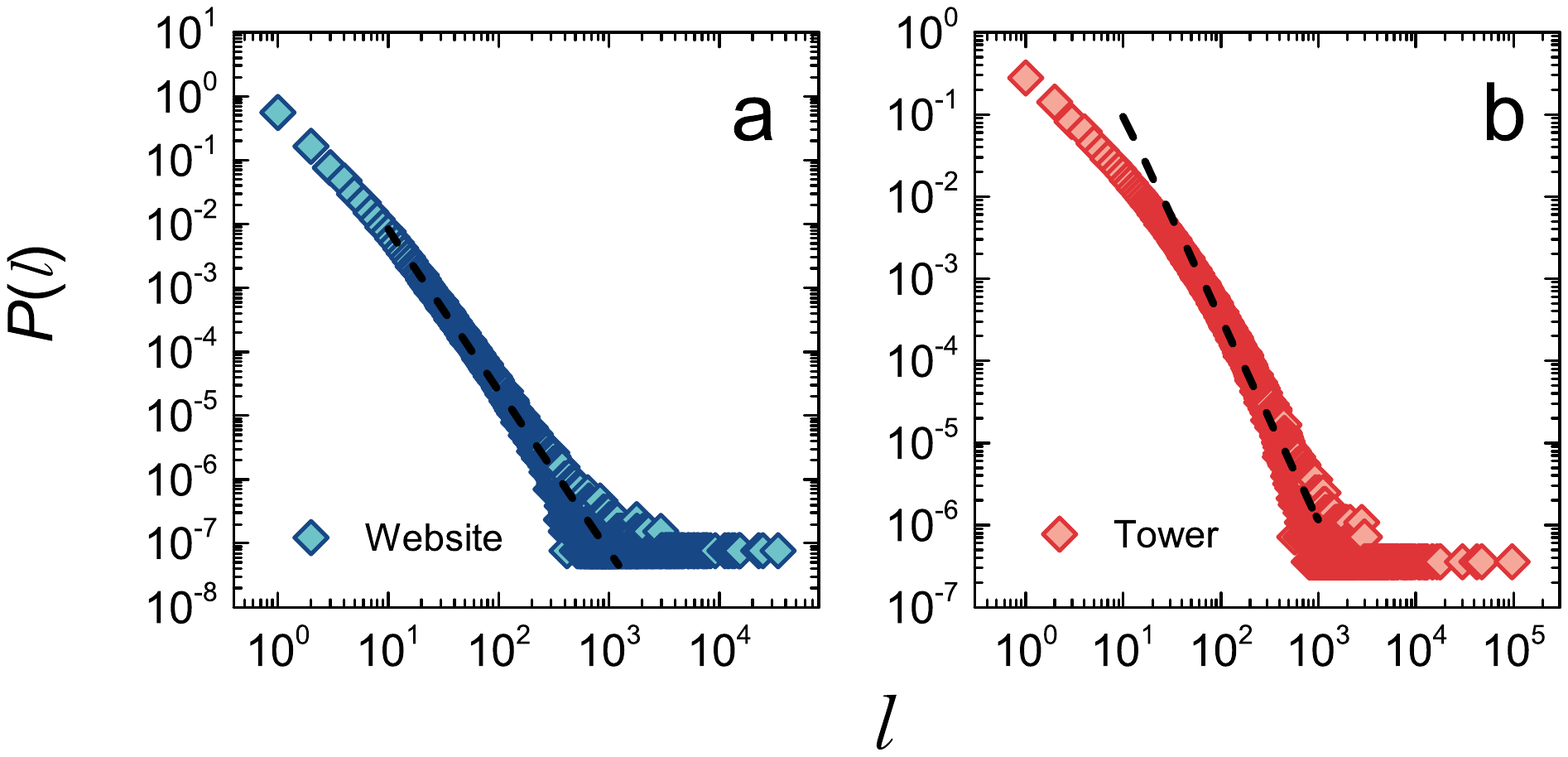,width=0.9\textwidth}
\end{center}
\caption{{\bf The probability distribution of dwelling time at collective level}.
By aggregating the dwelling time $l$ from all participants of whole system, we
show their probability distribution at collective level for online \textbf{(a)}
and offline \textbf{(b)} cases, respectively. Both of them can be fitted by
$P(l) \sim {l^{ - \alpha }}$ with the close exponent, i.e., $\alpha  = 2.51 \pm 0.02$
for online case and $\alpha  = 2.46 \pm 0.02$ for offline case.}
\label{fig:dwellcollective}
\end{figure*}

On the other hand, we also aggregate $l$ from all participants to quantify the non-Markovian character at collective level. As shown in Figs.~\ref{fig:dwellcollective}(a) and (b), it can be found that the distributions of $l$ for online and offline cases follow power-law form at multiply scales although a little deviation exists
at top scale. It is easy to understand that participant has inertia to stay at a particular node, resulting
in the power-law distributions of dwelling time. Additionally, at collective level, the values of $\alpha$ obtained through fitting the the power-law distributions for both online and offline cases at multiply scales are almost same, demonstrating the non-Markovian character generally roots in human mobility and suggesting the generic mechanism governing dynamics of human mobility no matter of its different behavior.

\begin{figure*}[!ht]
\begin{center}
\epsfig{figure=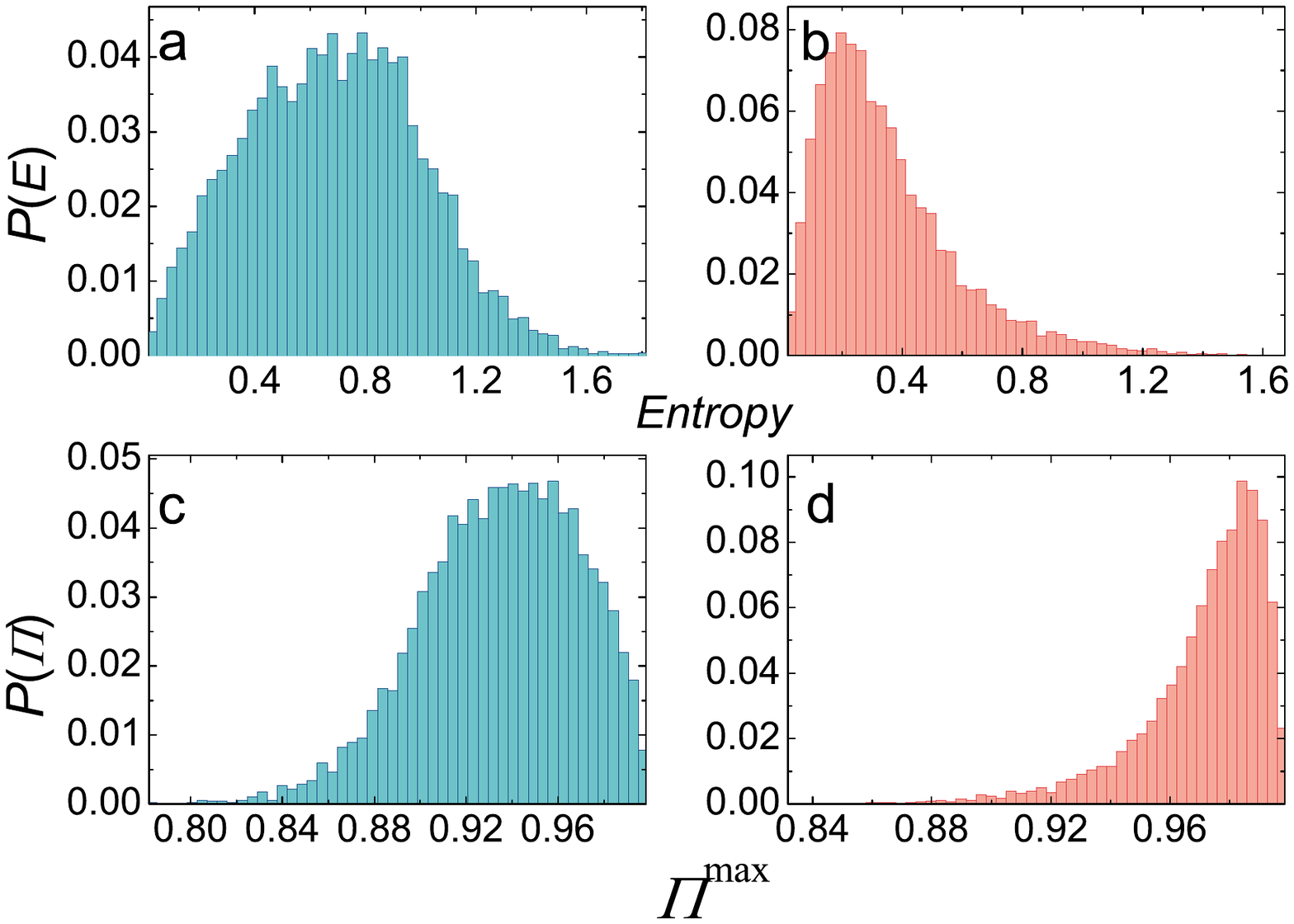,width=0.9\textwidth}
\end{center}
\caption{{\bf The probability distributions of individual entropy and predictability}.
For all participants of whole system, the probability distributions of entropy with the average $\langle E\rangle \simeq 0.69$
and $0.35$ for online \textbf{(a)} and offline \textbf{(b)} cases are presented to characterize the intense nonuniformity of system.
In turn, the probability distributions of corresponding predictability with
the average $\langle \Pi\rangle \simeq 0.94$ and $0.97$ for online \textbf{(c)}
and offline \textbf{(d)} cases confirm the lower entropy. Combined with the non-Markovian character in human mobility,
the higher predictability may be as a result of memory and non-randomness in dynamics of human mobility.}
\label{fig4entropypred}
\end{figure*}

\emph{Entropy and predictability}.
Entropy is the most fundamental quantity describing the uniformity of system and
the degree of predictability from time series~\cite{Song2010a}. A lower value of
entropy implies the intense non-uniformity of system and higher predictability, and
vice versa. Here, we discuss how to measure the entropy of sequence (i.e.,trajectories)
in human mobility. For this purpose, an estimator based on Lempel-Ziv data compression algorithm~\cite{Ziv1977,Kontoyiannis1998,Song2010a} is used to obtain the entropy of a given
sequence with $N_{a}$ actions as follows,
\begin{equation}
E \approx {(\frac{1}{N_{a}}\sum\limits_j {{\Lambda _j}} )^{ - 1}}\ln N_{a}.
\end{equation}
where $\Lambda _j$ is the length of the shortest substring starting at
position $j$ which doesn't previously appear from position $1$ to $j-1$.
Kontoyianis \emph{et al.}~\cite{Ziv1977,Kontoyiannis1998} have proved that $E$ converges
to the actual entropy as $N_{a}$ approaches infinity. We compute the values of $E$ over all
participants for both online and offline cases. Figure~\ref{fig4entropypred}(a) and (b) present
their distributions of entropy respectively, whose averages are different from
$\langle E \rangle \simeq 0.69$ for online case to $\langle E \rangle \simeq 0.35$ for offline case.
However, the low entropy are both observed in these two cases, suggesting higher predictability in human mobility.

Subjecting to Fano's inequality in~\cite{Fano1961}, the predictability ${\Pi ^{\max }}(E,S)$
when individual with entropy $E$ moves between $S$ nodes is determined by
\begin{equation}
E =  - [{\Pi ^{\max }}{\log _2}{\Pi ^{\max }} + (1 - {\Pi ^{\max }})
{\log _2}(1 - {\Pi ^{\max }})] + (1 - {\Pi ^{\max }}){\log _2}(S - 1).
\end{equation}
We compute ${\Pi ^{\max }}$ separately for each participant for online and offline cases,
and show their distributions shown in Figs.~\ref{fig4entropypred}(c) and (d). It can be
found that the higher predictability is presented for online and offline cases, and
$\langle \Pi\rangle$ is $0.94$ and $0.97$ respectively well agreeing with the lower entropy.
Combined with non-Markovian character in human mobility, we state that the high predictability
may be as a result of memory and non-randomness in dynamics of human mobility.

\subsection*{Model and Simulation}
\begin{figure*}[!ht]
\begin{center}
\epsfig{figure=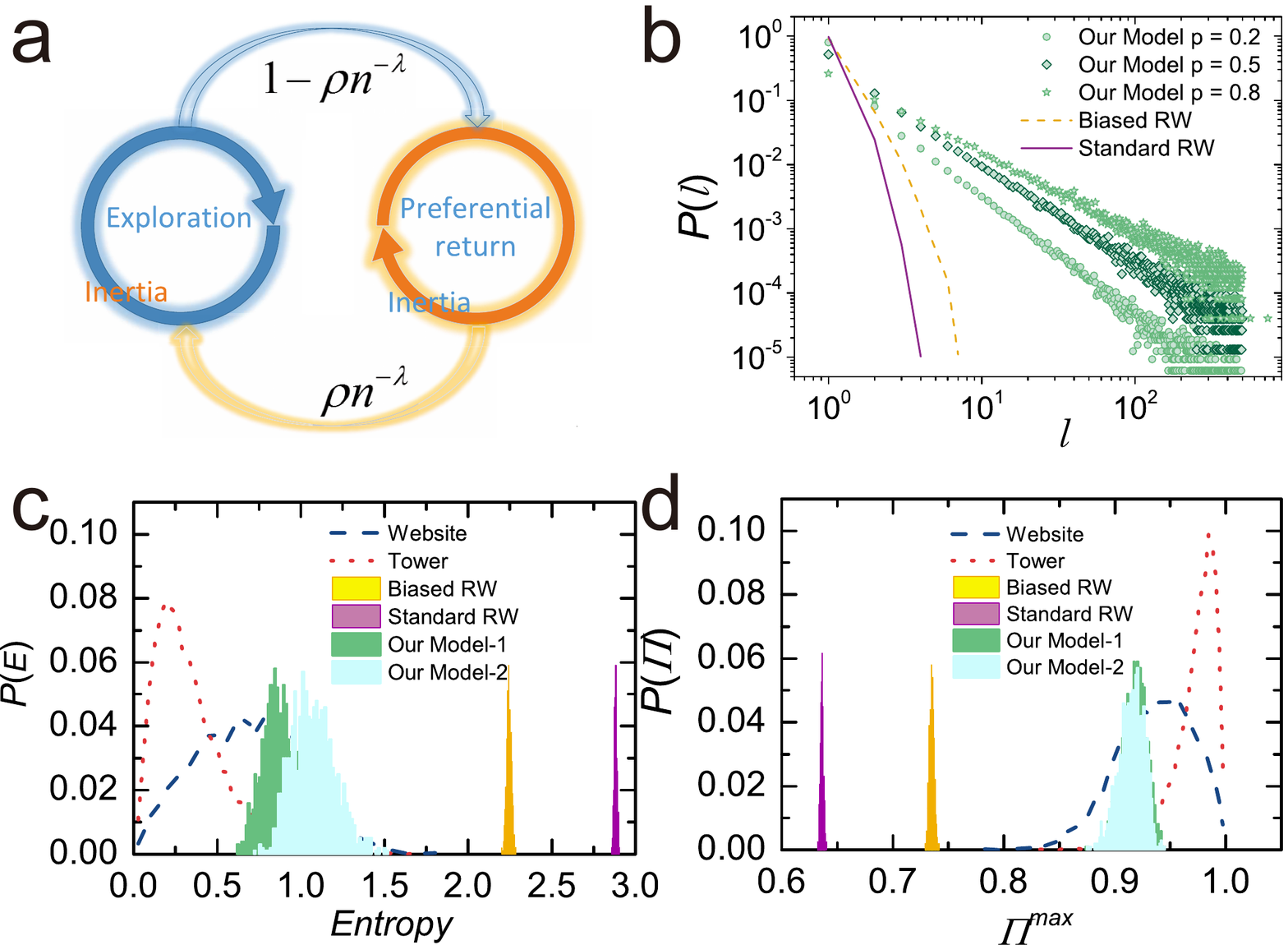,width=0.9\textwidth}
\end{center}
\caption{
\baselineskip=12pt
{\bf The schematic illustration of the proposed model of human mobility and the simulations}.
\textbf{(a)} Schematic illustration of the model, where an individual
can enter one of the two dynamically complementary states at each hopping
step: exploring a new node with the probability $\rho n^{-\lambda}$ (the
state of ``Exploration'') or returning preferentially to a previously visited
node with the probability $1 - \rho n^{-\lambda}$ (the state
of ``Preferential return''). Regardless of which state occurring,
a selected node is plugged by the inertial effect, a microscopic process
modeled by ERW with tunable parameter $p$.
These experimental results are obtained from $20, 000$ independent model simulations, each
using $10^6$ time steps for the parameter settings of $\{\rho, \lambda \}=\{0.49, 0.39\}$
and $\{0.51, 0.22\}$. \textbf{(b)} The robust power-law scaling behaviors of $P(l)$
restricted to the tunable parameter $p$ are obtained from the simulations of the proposed model,
in good consistency of the analytic result, $P(l) \sim l^{-(2-p)}$. And, the experimental results
of SRW and BRW are indicated by the solid purple line and orange dash line, which are far departure from
the empirical one. \textbf{(c)} and \textbf{(d)} The entropy and predictability of human mobility. The
empirical results for online and offline cases, corresponding to The blue dash and red dot
lines respectively, are used to guide the comparisons among models. As shown in Fig.\ref{fig:model}
(c) and (d), the experimental results of the proposed model, i.e., model-1 for online case
and model-2 for offline one, belong to the empirically statistical range, suggesting the
lower entropy and higher predictability in comparison with those of SRW and BRW.}
\label{fig:model}
\end{figure*}

The observed non-Markovian character as well as low entropy and high predictability infer that a participant visits to next
node is unlikely to be random and governed by underlying mechanism. Thus, it is meaningful to learn that why these scaling
laws emergence from human mobility and what is the underlying mechanism governing
the dynamics of human mobility with the low entropy and high predictability. Then, to acquire a reasonable mechanism characterizing
non-Markovian dynamics, three ingredients derived from empirical results, i.e., preferential return, exploration and inertia, are
employed to depict a protype model. A schematic illustration of proposed model is shown in Fig.~\ref{fig:model}(a). Concretely speaking,
a participant has two alternative choices, exploring a new node and revisit one of previous nodes. The probability,
making a choice of exploration, $\rho n^{-\lambda}$, is determined by statistical analysis of all participants,
whereas the complementary probability $1-\rho n^{-\lambda}$ is for another choice. Herein, the values of
$\rho$ and $\lambda$, estimated from real data sets, are 0.49 and 0.51 for online case and 0.39 and 0.22 for offline case
(see Fig.~S3 in Supplementary Information). When revisiting previous nodes, each of them has a probability of preferential return.
Therefore, an assumption, known in network science and human mobility~\cite{Barabasi1999,Song2010,Zhao2013b}, is that participants revisit one particular
node according to the prior visiting probability:
\begin{equation} \label{eq4}
\Pi = \frac{{{f _i}}}{{\sum\limits_{j = 1}^S {{f _j}} }},
\end{equation}
where $f _i$ is the frequency of visiting node $i$,
i.e., the accumulated dwelling time at node $i$. Once a particular node is chosen, she/he can keep staying at the selected node
for a time interval as a result of her/his inertia effect characterized by dwelling time. And the time interval can be simulated by
an excited random walk (ERW) with tunable parameter $p$~\cite{Antal2005}, that is, it equals to the time that
the walker \emph{first} returns to the starting point. As the walker eventually return to the starting point with probability 1 and
his first return time $t$ decays as $t^{-(2-p)}$, ERW associates with the non-Markovian character of the dwelling time.
Note that $p$ is nonnegative and usually less than 1.

For the comparison between the proposed model and those models with hypothesis of Markovian dynamics, we introduce two models,
standard random walk (SRW) visiting next node with uniform probability and biased random walk (BRW) visiting next node
subjecting to unequal chances. Based on all three models, the simulations are independently performed with $20,000$ agents,
each evolves $10^6$ time steps. The numerical results are shown in Fig.~\ref{fig:model}(b)-(d). On one hand,
the scaling law predicted the proposed model with different $p$ are much closer to empirical result comparing with SRW and BRW, although
its values are a little departure from these empirical ones. (see in Fig.~\ref{fig:model}(b)). On the other hand, the lower entropy
and higher predictability obtained from the proposed model well belong to the ranges observed from all participants, while those of SRW and
BRW totally deviate from these ranges. These results suggesting these two Markovian models are not able to perfectly describe
human mobility for both online and offline cases.

\section*{Discussion}
Despite making large advances in human dynamics, especially human mobility~\cite{Barabasi2005,Oliveira2005,Dezso2006,Zhou2008,
Goncalves2008,Wu2010,Gonzalez2008,Song2010,Song2010a,Simini2012,Yan2013}, we still
pay much attention on its dynamic characters because of the obscurity of generical
mechanism. Through analyzing a massive data set including both online and offline
trajectories of human mobility, the nonuniform dynamic process of human mobility
is shown by the heterogenous rank distribution of nodes of visual network, and
its non-Markovian character is suggested by the emergence of
scaling laws in distributions of dwelling time at both individual and collective
levels. The non-Markovian character implies the memory in dynamics of human mobility,
leading to the low entropy and high predictability. Furthermore, a protype model involved with three ingredients, preferential return, exploration and inertia is introduced to reproduce transition process of human mobility and account for the origin of empirical results. The simulations show that the proposed model presents a better prediction of scaling law comparing with those models with Markovian dynamics, and behaves the lower entropy and higher predictability close to statistical average for all participants. Note that the present model may be
so simple as to be not able to unveil extra dynamic characters but enough to figure out the non-Markovian characters.

The scaling laws uncovered from data and the protype model developed
accordingly can be applied to addressing significant problems ranging
from human-behavior prediction and the design of search
algorithms~\cite{Vespignani2009,Song2010} to controlling spreading
processes~\cite{Balcan2011,Zhao2012}. Moreover, the non-Markovian character of human mobility could make
to deeply understand other studies such as synchronization~\cite{Arenas2008} and temporal networks~\cite{Holme2012}. As a demonstration, we have probed the degree of predictability of human behavior and its origin.

\section*{Materials and Methods}

\subsection*{Data sets description.}
The massive data set used is from a large-scale real communication system based mobile termination, where
participant browses websites when they are moving from one place to another. Thus, it includes two groups of
records, online trajectory of browsing weibsites and offline trajectory of visiting towers.
In order to improve the quality of trajectory reconstruction, we remain these participants whose trajectory achieves at least 20 distinct websites and towers, and the average visiting times to each website or tower is more than 10. Finally, the total number of participants is $8,929$.

\subsection*{Maximum likelihood estimation and Kolmogorov-Smirnov test.}
Maximum Likelihood Estimation is a solid tool for learning parameters as a data mining model. Its principle makes the desired probability distribution be the best fit of observed data, and its methodology tries to do two things. First, it is a reasonably well-principled way to work out what computation you should be doing when you want to learn some kinds of model from data. Second, it is often fairly computationally tractable. The details of the maximum likelihood method have been widely published. Herein, we perform such fits according to~\cite{Clauset2009}.

In statistics, $Kolmogorov-Smirnov$ (KS ) test is a nonparametric methodology that compares either an observed distribution, $S(x)$, with a theoretical distribution, ${F^*}(x)$ , or two observed distributions. In either case, the procedure is involved with forming the cumulative distributions of $S(x)$ and ${F^*}(x)$ and finding the size of the largest difference between them. The KS test is based on the following test statistic:
\begin{equation}
D = \sup |{F^{*}_{c}}(x) - S_{c}(x)|
\end{equation}
where smaller $D$ values correspond to better fit with theoretical distribution.

\subsection*{Standard, biased and excited random walks}
For standard Random Walk (SRW), $S$ distinct nodes are distributed and then the walker
select one node to visit randomly (with probability $1/S$) at each step.
The process is repeated until all of the $S$ nodes are visited. While for biased random
walk (BRW), the walker select one node to visit at each step according to the preferential
probability ${\Pi _i} = {{{f_i}} \mathord{\left/{\vphantom {{{f_i}}
{\sum {{f_i}} }}} \right.\kern-\nulldelimiterspace} {\sum {{f_j}} }}$
depending on its visit frequency~\cite{Barabasi1999,Song2010}. The large difference
between SRW and BRW is that SRW can bring larger entropy and lower predictability than BRW.
However, both of them cannot simultaneously include the low entropy and high predictability
observed from empirical data.

The excited random walk (ERW) on one dimension represents a simple and surprisingly rich example
of a non-Markov process and possesses a basic feature that transitory nature of the bias~\cite{Antal2005}.
For any $p<1$, it is recurrent, that is, the walker revisits sites unlimited and
the probability to eventually return to the starting point equals to 1. Its process on the infinite
chain of sites is simply described as: the walker, starting from a certain site to first visit,
hops to the right with probability $p$ and to the left with complementary probability $1-p$; then
if the walker hops to a previously visited site, at next time step the hopping probabilities to
the left and right are both $1/2$, otherwise they are biased; the walker finally stops hopping
until he first return to the starting site. The time steps across a whole ERW are deemed as the time interval
of inertial effect.

\section*{Acknowledgments}
This work is jointly supported by the NNSFC (Grant Nos.91024026 and 61004102), Special
Project of Sichuan Youth Science and Technology Innovation Research Team (No.2013TD0006),
and the Fundamental Research Funds for the Central Universities (Grant Nos.ZYGX2011YB024 and ZYGX2012J075).
SMC acknowledgments the financial support of the Open Foundation of State key Laboratory of Networking
and Switching Technology (Beijing University of Posts and Telecommunications) (Grant No.SKLNST-2013-1-18) and the startup fund of UESTC.

\clearpage

\clearpage

\section*{Figure Legends}

\end{document}